\documentclass[aps,rmp,reprint]{revtex4-1}

\usepackage{graphicx}
\usepackage{dcolumn}
\usepackage{bm}

\date{September 30, 2017}

\begin{document}

\title{A Review of Laser-Plasma Ion Acceleration}
\author{A.Macchi}
\affiliation{National Institute of Optics, National Research Council (CNR/INO),  Adriano Gozzini laboratory, via Giuseppe Moruzzi 1, 56124 Pisa, Italy}\email{andrea.macchi@ino.cnr.it}\homepage{http://www.andreamacchi.eu}
\affiliation{Enrico Fermi Department of Physics, University of Pisa, largo Bruno Pontecorvo 3, 56127 Pisa, Italy}

\begin{abstract}
An overview of research on laser-plasma based acceleration of ions is given. The experimental state of the art is summarized and recent progress is discussed. The basic acceleration processes are briefly reviewed with an outlook on hybrid mechanisms and novel concepts. Finally, we put focus on the development of engineered targets for enhanced acceleration and of all-optical methods for beam post-acceleration and control.
\end{abstract}

\maketitle
\tableofcontents

\section{Introduction}
\label{sec:intro}
The observation of intense multi--MeV proton emission from solid targets irradiated at ultra-high intensities in three independent experiments \citep{clarkPRL00,maksimchukPRL00,snavelyPRL00} performed in the year 2000 promptly boosted a great research effort on laser-driven (or laser-plasma) ion accelerators. Such research has been oriented to several foreseen applications in nuclear fusion, medicine and high energy density science. Moreover, it has represented a major motivation for the development of laser systems with increasing peak power, towards the multi-petawatt (PW) frontier \citep{dansonHPL15}.
 
\begin{figure}[t!]
\begin{center}
\includegraphics[width=0.45\textwidth]{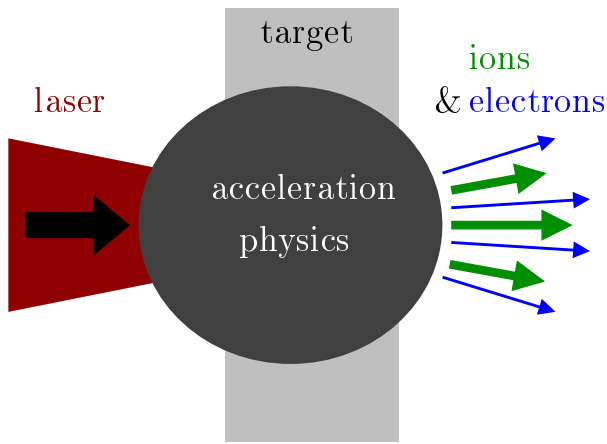}
\end{center}
\caption{Schematic rappresentation of laser-driven (or laser-plasma) acceleration of ions, showing the essential elements: the laser pulse, the target, and the ion beam (charge-neutralized by comoving electrons). All the physics which leads to conversion of the energy and momentum of the laser pulse into energy and momentum of the ions is concealed in the central ``black box''.}
\label{fig:basicLIA}
\end{figure}

Fig.\ref{fig:basicLIA} shows a very schematic representation of the experimental scenario: a high intensity laser pulse is focused on one side of the target (commonly referred to as the ``front'' side) and a bunch of energetic protons\footnote{Proton emission was also observed using targets with no hydrogen in their chemical compositions (e.g. metallic targets). This is because hydrogen-containing impurities (water, hydrocarbons) are typically present on the target surface.} (or heavier ions) is detected on the opposite (``rear'') side. The ``black box'' in Fig.\ref{fig:basicLIA} conceals a complex acceleration physics, involving several possible mechanisms.

This article aims to give a concise overview on the motivations, principles, state-of-the-art, physics and perspectives of laser-driven ion acceleration. The goal is to provide an introduction to this field accessible to the non-specialist reader and to summarize major achievements and most recent developments. More complete and detailed presentations may be found in several review papers both by the present author and coworkers \citep{macchiRMP13,macchiPPCF13,macchi-borghesi} and by others \citep{daidoRPP12,fernandezNF14,schreiberRSI16}. 

\subsection{Properties of Laser-Accelerated Protons}
\label{sec:properties}
The reason for a sudden, enormous interest in laser-accelerated protons has been in their foreseen use in several applications. Common to these latter is the exploitation of the strongly localized energy deposition (the Bragg peak) of high energy ions \citep{knoll-book,ziegler-book08} combined with the unique properties which were apparent in the early experiments with intense lasers. Such properties include high laser-to-proton energy conversion efficiency ($>10\%$ estimated for a petawatt driver \citep{snavelyPRL00}) resulting in up to a few tens of joules total energy of the protons; the very large number of protons per shot (up to $\sim 10^{13}$); the proton collimation within a typical cone angle of few tens of degrees; the focusability of the protons by a simple shaping of the target rear surface; the very low emittance\footnote{The emittance needs to be properly defined and measured for non-monoenergetic beams \citep{nuernbergRSI09}.} (down to $\sim 10^{-3}$~mm~mrad \citep{cowanPRL04,nuernbergRSI09}); the inferred ultrashort duration in the ps ($10^{-12}$~s) regime. In addition, the protons are accompanied by a co-moving electron cloud, which ensures charge neutralization and prevents the proton bunch from exploding due to electrostatic repulsion (``Coulomb explosion''), so that one could even consider this to be a quasi-neutral ``plasma bunch''.

Some of the differences with the protons delivered by standard accelerators (such as the compact cyclotrons, synchrotrons and linacs used in medical physics or other applications) appear to be very remarkable. The accelerating gradient, i.e. the typical value of the electric field on ions, is of the order of MV/$\mu$m which is some four orders of magnitude greater than the value in a linac (up to 100~MV/m).  
A pulse duration in the picosecond range is at least some three orders of magnitude shorter than what is achievable with standard accelerators, which are in the range between ns and $\mu$s. Combined with the high number of particles per bunch, this yields a peak current of several kA (or even higher), which may be compared with the mA steady current in a cyclotron. Finally, the transverse emittance is typically three orders of magnitude lower than the typical value in a linac ($\sim$~1~mm~mrad), and the longitudinal emittance is very low as well. 

The above mentioned properties make laser-accelerated protons highly promising for any application requiring an extremely localized (both in space and time) energy deposition in dense matter. The first observations were promptly followed by the proposal of using laser-accelerated protons to create a hot spot in an Inertial Confinement Fusion (ICF) target \citep{rothPRL01,ruhlPPR01,atzeniNF02}, thus providing an alternative ignitor for the fast ignition concept \citep{tabakPoP94} in ICF. Soon after, laser-accelerated protons were proposed for oncological ion beam therapy (IBT) as a possibly more compact and cheaper option than that using conventional accelerators \citep{bulanovPLA02,bulanovPPR02,fourkalPMB03,malkaMP04}. As another medical application, production of short-lived radioisotopes was also proposed and preliminary investigations were performed \citep{nemotoAPL01,santalaAPL01,spencerNIMB01,fritzlerAPL03,ledinghamS03}.

The low emittance favored the application of laser-accelerated protons for radiography and imaging \citep{rothPRSTAB02,cobbleJAP02,borghesiPRL04}. In particular, the broad energy spectrum combined with the short duration of the proton bunch has enabled the development of single-shot detection of electromagnetic fields in laser-plasma interaction phenomena with picosecond temporal resolution \citep{borghesiPoP02,mackinnonRSI04}. This innovative application has yielded much valuable information on the nonlinear dynamics of plasmas, including the proton acceleration mechanism itself \citep{romagnaniPRL05}.

\subsection{Principles of Ion Acceleration}
\label{sec:principles}
The special properties of laser-accelerated proton bunches have their roots in the \emph{coherent} (in the sense of \emph{collective} or \emph{cooperative}) nature of the acceleration process, which is basically different from conventional technologies. The coherent acceleration paradigm was outlined by 
\citet{vekslerSJAE57}, 
before the invention of the laser. Key features of this paradigm include: 1) the accelerating field on each particle is proportional to the number of accelerated particles (the larger their number, the higher the kinetic energy of the single particle); 2) the field is localized in space and synchronized in time with the accelerated particles; 3) eventually the acceleration process produces globally charge-neutralized bunches. These latter three features are realized in the acceleration of ions occuring via the interaction with sufficiently dense targets, i.e. in \emph{laser-plasma} acceleration.

While collective plasma dynamics is the basis of the unique properties and potential applications of laser-accelerated ion bunches, its complex nonlinearity poses great challenges of control, stability, and modeling with respect to traditional approaches. The basic acceleration mechanism underlying most of the experiments reported so far, commonly named Target Normal Sheath Acceleration (TNSA), has reached a good level of reliability and robustness, and provides a framework for further developments such as ``all-optical'' (i.e. laser-controlled) bunch control and post-acceleration. However, foreseen applications have stringent requirements on properties such as the energy per particle, the spectral distribution, and the suitability for high-repetition rate operation, as well as others. The parameters and characteristics of laser-accelerated protons have still to reach such requirements, and it is still uncertain whether the availability of multi-PW lasers in the next few years combined with developments of innovative targets will be sufficient for such aims, at least for what concerns TNSA-based approaches.
This issue has stimulated the proposal of alternative schemes, such as e.g. Radiation Pressure Acceleration (RPA) or Collisionless Shock Acceleration (CSA) whose investigation is still relatively preliminary compared to that for TNSA. 
The physics of such mechanisms will be discussed in sec.\ref{sec:physics}.

\section{State of the Art}
\label{sec:state}
Reporting on the state-of-the-art in laser-driven ion acceleration is not straightforward for several reasons. For instance, progress has been achieved on several properties (e.g. the maximum proton energy, the conversion efficiency, the spectral width \ldots) but in different experiments and, in most cases, for different acceleration mechanisms (section \ref{sec:physics}). Moreover, the experimental characterization of laser-driven ion bunches is not easy. Established methods and instruments have required modifications and adaption, and new diagnostic techniques have been developed (for reviews see, e.g., \citet{boltonPM14} and section II.E of \citet{macchiRMP13}). While such effort has allowed a study of laser-driven ion acceleration over a wide range of laser parameters (intensity, energy, duration, polarization, \ldots), not all the relevant properties of the accelerated ions are usually measured in a single experiment. In addition, a precise control of the experimental conditions is also challenging with high-power lasers, and this may account for variations observed between experiments performed in conditions which would seem similar at a first glance. A likely consequence of these issues is that the scaling of the most important characteristics (such as the energy per particle) with laser and target parameters is still unclear to a large extent, despite the large number of investigations performed. The reader should keep in mind these issues and remarks in the following description of experimental achievements.
 
Most of the experiments performed so far have dealt with the acceleration of protons from solid targets. In the literature, progress in the field has been mostly monitored and claimed on the basis of the maximum proton energy observed, since for instance reaching the energy window for IBT applications (60--250~MeV, where at least 150~MeV is required for deeply seated tumours) has been considered to be a major goal. The energy spectra of protons are typically broad and in most-cases exponential-like, and the maximum energy corresponds to the upper cut-off in such spectra. However, the observed cut-off in the spectrum may depend on the sensitivity of the detector used and on the level of background noise, so that it is not always clear how precisely a maximum energy can be measured, and to what extent the value is affected by diagnostic factors. A comparison of absolute differential spectra, obtained with calibrated detectors yielding the number of particles per energy interval, can be (when such spectra are available) less prone to diagnostic factors and thus give a safer indicator of the acceleration performance. Moreover, the number of protons per energy slice (and possibly per opening angle) is an important parameter for foreseen applications which require a sufficient particle flux in addition to a given energy range. 

\subsection{Progress in Proton Energy Enhancement}
\label{sec:longpulse}

\begin{figure}[t!]
\begin{center}
\includegraphics[width=0.45\textwidth]{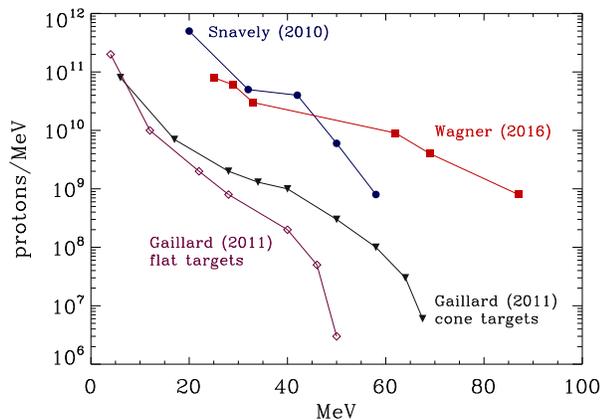}
\caption{Proton energy spectra from the three experiments reporting new world records in the cut-off energy from 2000 to 2016: data were taken from 
\citet{snavelyPRL00}, 
\citet{gaillardPoP11} 
and \citet{wagnerPRL16}. For the experiment of 
\citet{gaillardPoP11} 
the record energy was achieved using specially shaped ``cone'' targets; the spectrum obtained with standard ``flat'' targets is reported for comparison. See text for parameters and details.}
\label{fig:psdata}
\end{center}
\end{figure}

Figure~\ref{fig:psdata} shows a comparison of calibrated spectra from three experiments \citep{snavelyPRL00,gaillardPoP11,wagnerPRL16} which established new world records for the cut-off energy at the time of the publication of the results in the period between 2000 and 2016. All these three experiments were performed with ``large'' laser systems, delivering several tens of joules onto a solid target, with a typical pulse duration of several hundreds of femtoseconds. The facilities where the experiments were performed and the corresponding reported values of the energy ``on target'' (i.e. contained into the focal spot) and the pulse duration were, respectively, the petawatt system at Lawrence Livermore National Laboratory (LLNL) in the year 2000 (150~J, 500~fs) \citep{snavelyPRL00}, the TRIDENT laser at Los Alamos National Laboratory (LANL) in 2011 (39$\pm$7~J, 670$\pm$130~fs) \citep{gaillardPoP11}, and the PHELIX laser at GSI Helmholtzzentrum fuer Schwerionenforschung Darmstadt in 2016 (48-60~J, 500-800~fs) \citep{wagnerPRL16}. 
The corresponding values of the conversion efficiency of laser energy into proton energy were 12\% (LLNL) obtained integrating the proton spectrum for energies $>$10~MeV, 1.75\% (LANL) and 7\%$\pm$3\% (GSI) both for proton energies $>4$~MeV. The LLNL and GSI data were both obtained with plastic targets (CH and CH$_2$, respectively) but with very different values of the target thickness (100~$\mu$m and 0.9~$\mu$m, respectively). The LANL data were obtained with copper (Cu) targets with 10~$\mu$m thickness, and the highest energies were obtained with special targets where the laser was focused into a microcone placed at the front side (i.e. the target side exposed to the laser irradiation), obtaining an increase to 67.5~MeV as compared to 50~MeV from flat targets (see Fig.~\ref{fig:psdata}). 

Comparing the spectra from the three experiments in Fig.~\ref{fig:psdata} suggests that the cut-off increase obtained at LANL with respect to LLNL data may be effectively due to an increased sensitivity of the detector, since the number of protons at the cut-off is smaller by about two orders of magnitude. However, it is remarkable that the cone targets yield similar proton energies with much less laser energy. 
The reduction of the target thickness in the GSI experiment appears to produce a substantial progress with respect to the LLNL data (also at considerably less energy), since the energy cut-off is increased by nearly 30~MeV at almost the same number of particles. 

The above analysis also indicates that the cut-off energy, apart from the difficulties related to its definition and measurement, is not the only parameter by which the performance of ion acceleration should be measured: an increase of only 30~MeV over 16 years might appear as a slow progress, but one should consider that, while the full energy available with large systems delivering picosecond pulses has not increased during this time period, significantly less energy has been used to obtain similar or even higher proton energies. 

One may also argue that the large systems, producing hundreds of Joule pulses, used for the experiments reported in Fig.\ref {fig:psdata} are far from being compact and generally unsuitable for high repetition rate operation, which is a key requirement for most applications. For this reason, it is of interest to evaluate the progress obtained with ``smaller'' laser systems which may operate at $10-10^3$~Hz rate and typically have a pulse duration of a few tens of fs and an available energy $\lesssim$10~J.

\subsection{Proton energy scaling with short-pulse drivers}
\label{sec:shortpulse}
In reviewing data obtained with smaller short-pulse systems we also select experiments where a calibrated energy spectrum is available. 
In addition, as an additional criterion we consider only experiments using ``high-contrast'' pulses. High power laser systems typically do not produce ``clean'', isolated short pulses; indeed, the ``main'' pulse of sub-picosecond duration is preceded by other pulses of the same duration and lower power, a few nanoseconds ``pedestal'' pulse, and another pedestal of picosecond duration produced by imperfect recompression of the main pulse.
When aiming at the highest intensity of the main short pulse, the prepulses may be already intense enough to cause ionization and heating in the target, producing a ``preplasma'' at the interaction surface. 
A controlled short-pulse interaction requires pulses with a sufficiently high ``contrast'' ratio between the intensities of the main pulse and the prepulse(s) must be high enough to prevent target damage and preplasma formation, to which the laser-plasma coupling is highly sensitive.
In recent years the development of optical devices such as the plasma mirror \citep{dromeyRSI04,thauryNP07} made possible achieving pulse contrast values of $10^{10}$ (typically measured a few ps before the short fs pulse) and beyond. This means that even at the highest short pulse intensities $\sim 10^{21}~\mbox{W cm}^{-2}$ available today, the prepulse intensity is $\sim 10^{11}~\mbox{W cm}^{-2}$, which is under the ionization threshold of most target materials \citep{vonderLindeJOSAB96}. 
It is worth noting, however, that our choice to consider high-contrast experiments is only to have similar interaction conditions, and it does not imply that high contrast always favors the enhancement of the proton energy or other properties.

The selected experiments were performed in different laboratories using laser systems of different nominal power, 
i.e. 
the 3~TW laser at I3M Valencia \citep{seimetzELI06},
the LLC 30~TW laser in Lund \citep{neelyAPL06}, 
the DRACO 150~TW laser at HZDR Dresden \citep{zeilNJP10,zeilNC12,zeilPPCF14},
the LiFSA 100~TW \citep{choiAPL11,margaronePRL12} and PULSER 1~PW \citep{kimPRL13,margaronePRSTAB15,passoniPRAB16} lasers at GIST Gwangju, 
the JKAREN 200~TW laser at JAEA/KPSI Kyoto \citep{oguraOL12}, 
and the GEMINI 200~TW laser at RAL \citep{greenAPL14}.
The corresponding ranges of pulse duration ($\tau_L$) and intensity on target ($I_L$) are $25-40$~fs and $4 \times 10^{18}-2 \times 10^{21}~\mbox{W cm}^{-2}$. In order to reduce possible uncertainties due to different optical transport and focusing systems, for each experiment the energy \emph{on target}, contained within the full-width-half-maximum (FWHM) of the intensity distribution in the focal spot has been considered. The targets have different thicknesses in the range $0.01-4.0~\mu\mbox{m}$. To simplify the analysis, we only consider the higher energy tail of each experimental spectrum, which we found to be satisfactorily approximated by a simple exponential function $N_{\mbox{\tiny p}}({\cal E})=N_{\mbox{\tiny p0}}\exp(-{\cal E}/T_{\mbox{\tiny p}})$ with ${\cal E} \leq {\cal E}_{\mbox{\tiny co}}$, the cut-off energy. The ``temperature'' $T_{\mbox{\tiny p}}$ is a parameter giving information on the mean proton energy and the spectral roll-off with increasing ${\cal E}$.

\begin{figure*}[btp]
\begin{center}
\includegraphics[width=\textwidth]{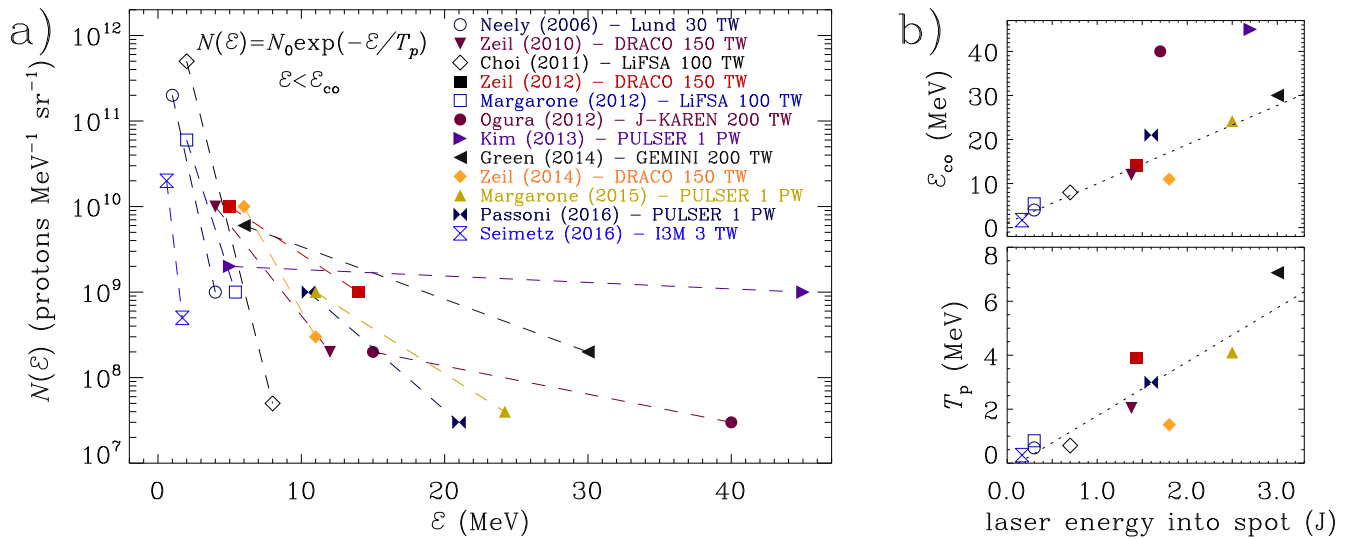}
\caption{Frame a): proton energy spectra from experiments using high contrast, sub-100~fs, sub-10~J laser pulses and thin solid targets, shown as simple exponential interpolations (dashed lines) $N_{\mbox{\tiny p}}({\cal E})=N_{\mbox{\tiny p0}}\exp(-{\cal E}/T_{\mbox{\tiny p}})$ (with ${\cal E} \leq {\cal E}_{\mbox{\tiny co}}$, the cut-off energy) of the high-energy tail of experimentally measured spectra. 
Frame b): the parameters ${\cal E}_{\mbox{\tiny co}}$ (top) and $T_{\mbox{\tiny p}}$ (bottom) as a function of the laser pulse energy in the focal spot. 
Data are from 
\citet{kimPRL13,passoniPRAB16,seimetzELI06,neelyAPL06,zeilNJP10,zeilNC12,zeilPPCF14,choiAPL11,margaronePRL12,oguraOL12,greenAPL14,margaronePRSTAB15} 
as indicated in the text.
Empty and filled symbols are for intensities in the $I_L=(0.4-5)\times 10^{19}~\mbox{W cm}^{-2}$ and $I_L=(0.8-2)\times 10^{21}~\mbox{W cm}^{-2}$ ranges, respectively. The pulse durations are in the $\tau_L=(25-40)~\mbox{fs}$ range. All the targets are planar foils of various thickness (in the $0.05-4.0~\mu\mbox{m}$ range).
}\label{fig:fsdata}
\end{center}
\end{figure*}

Fig.\ref{fig:fsdata} shows that the results of \citet{kimPRL13} and \citet{oguraOL12} are at variance with other experiments for what concerns the proton spectral density $N_{\mbox{\tiny p}}$, the cut--off energy ${\cal E}_{\mbox{\tiny co}}$ and the ``temperature'' $T_{\mbox{\tiny p}}$ (out of scale in the plot). The rest of the data show an increasing trend for these parameters as a function of the energy on target $U_L$, with an apparent almost linear scaling of ${\cal E}_{\mbox{\tiny co}}=kU_L$ with $k \simeq 8.6~\mbox{MeV}/\mbox{J}$. 
Such linear scaling is similar to that proposed by \citet{zeilNJP10} on the basis of a parametric investigation with the DRACO laser and extends the latter to higher values of the pulse energy. The linear scaling is different from, and more favorable than the intensity dependence ${\cal E}_{\mbox{\tiny co}} \sim I_L^{1/2}$ which has been inferred in previous reviews from data obtained using short pulse lasers \citep{fuchsNP06,borghesiPPCF08,daidoRPP12,kieferPRE13}. Also $T_{\mbox{\tiny p}}$ shows a scaling which is almost linear with the energy in the focal spot. 
It is worth stressing that, although previous reviews of energy scaling vs. laser intensity or energy use a larger set of data than in Fig.\ref{fig:fsdata}, the scatter with respect to the proposed scalings appear to be larger than for the data selected in Fig.\ref{fig:fsdata} (also note that, unlike Fig.\ref{fig:fsdata}~b), data are most often represented on a logarithmic scale). The residual data scatter in Fig.\ref{fig:fsdata} may be ascribed to the variation of other parameters (laser pulse duration, target thickness and material, incidence angle, \ldots). The scaling of proton energy with such parameters is less clear and is not discussed here.
Some comments on the ``anomalous'' data in Fig.\ref{fig:fsdata} \citep{kimPRL13,oguraOL12} will be given at the end of section~\ref{sec:conclusions}, on the basis of the discussion of the acceleration mechanism in sec.\ref{sec:physics}.

\subsection{Progress in Ion Bunch Properties}
\label{sec:beamproperties}
The experimental effort for characterization and optimization of ion bunch properties different from the maximum energy per particle has been less extended and systematic. It is worth noting that optimization of the bunch properties needs to be guided by an understanding of the acceleration process. This is particularly needed for aims such as obtaining a narrow energy spectrum, which is \emph{qualitatively} different from the data shown in the preceding sections. Progress on this side has been mostly obtained via the engineering of target structure and chemical composition (see e.g. \citet{hegelichN06,schwoererN06,pfotenhauerNJP08}) or via the exploration of advanced acceleration mechanisms (see e.g. \citet{palmerPRL11,haberbergerNP12,karPRL12,palaniyappanNC15}). In most of these cases the results are either preliminary or need further confirmation. For instance a spectral width as low as 1\% has been measured using CO$_2$ lasers and gas jet targets \citep{haberbergerNP12} but the results are at variance with others obtained in apparently similar conditions (see sec.\ref{sec:CSA}). Perspectives will be discussed in sec.\ref{sec:conclusions} after the acceleration physics has been discussed.

For the case of broad energy spectra which are observed in most experiments, the bunch properties may be only defined with reference to a given energy range. Here we only mention recent progress related to characterization and optimization of selected properties such as ion bunch duration and conversion efficiency. More complete lists of achieved values can be found in the review papers cited in sec.\ref{sec:intro}. 

One of the most peculiar (possibly unique) properties is the ultrashort duration of the ion bunch. In early experiments, a picosecond duration at the source has been inferred and roughly estimated from the analysis of proton probing experiments \citep{borghesiPoP02,mackinnonRSI04}. However, until recently no direct measurements were reported. Of course the measurement of bunch duration must be associated not only to a well-defined energy band but also to a value of the distance from the source, since velocity dispersion will cause the bunch duration to increase along the propagation path. In recent experiments with the TARANIS laser ($\sim 600$~fs pulse duration) at the Queen's University of Belfast, the bunch duration has been measured via observations of proton-generated ionization in SiO$_2$ \citep{dromeyNC16}. For a bunch of protons with energy of $10\pm 0.5$~MeV, the effective duration measured in a SiO$2$ sample at a distance of $\sim 5$~mm was $3.5 \pm 0.7$~ps. This value opens up perspectives for ultrafast studies of ion-induced damage and energy deposition in dense matter. It might be possible to generate even shorter ion bunches by employing a laser driver with few tens of fs duration and specifically tailored bunch modulators.

The $\sim 12\%$ conversion efficiency of laser energy into protons in the 10-58~MeV range observed at LLNL in 2000 \citep{snavelyPRL00} has been an even more long-lasting record than the cut-off energy. An experiment performed on VULCAN in 2014 \citep{brennerAPL14} has obtained a $\sim 15\%$ efficiency for protons in the 3-30~MeV energy range. In order to optimize the acceleration process the experiment used a controlled short prepulse which was shown in previous experiments to enhance the cut-off energy and conversion efficiency \citep{markeyPRL10} and to produce modulations in the energy spectrum \citep{dollarPRL11}. 
These experiments provide an example of all-optical techniques for manipulation and optimization of proton acceleration, which will be discussed in sec.\ref{sec:postacceleration}.

\section{Ion Acceleration Physics}
\label{sec:physics}
The present section is devoted to a basic description of laser-plasma interactions at high intensities and of the main ion acceleration mechanisms\footnote{A more detailed tutorial introduction to the acceleration physics at a (mild) postgraduate level can be found in Chap.5 of the author's textbook \citep{macchi-book}.}, whose understanding is essential for developments of laser-driven ion sources. For each mechanism, most significant experimental confirmations are also mentioned.

\subsection{Laser-Plasma Interaction Scenario}
\label{sec:scenario}
In the interaction regimes of relevance to the present context, the laser pulse is intense enough to ionize matter almost instantaneously, and couples with the freed electrons which absorb energy and momentum from the electromagnetic field. The interaction typically leads to the generation of suprathermal electrons of high energy; such ``fast'' electrons tend to escape from the target generating regions of charge separation and related electrostatic fields in the presence of density gradients, particularly at target boundaries (``sheath'' regions), and in turn the electrostatic fields accelerate ions and drive the expansion of the plasma. Momentum absorption occurs due to a secular {ponderomotive force} (corresponding to the local flow of electromagnetic momentum) which modifies the electron density and consequently the electrostatic fields, leading as well to radiation pressure action on the plasma. Under suitable conditions, the combination of heating and {radiation pressure} can drive nonlinear shock waves which also lead to ion acceleration. The basic mechanisms we briefly describe below originate from the dominance of each of these effects, which however may  generally coexist in experiments leading to a complex acceleration scenario. 

In the following, we also emphasize laser and target requirements and developments needed to advance each mechanism. In order to characterize the interaction regime, two dimensionless parameters are particularly useful and important. The first one is the ratio between the free electron density $n_e$ in the target and the cut-off or ``critical'' density $n_c$ , i.e. the maximum value of the electron density above which the laser pulse does not propagate:
\begin{equation}
n_c=n_c(\omega_L)=\frac{m_e\omega_L^2}{4\pi e^2} \; ,
\label{eq:nc}
\end{equation}
where $\omega_L$ is the laser frequency. Eq.(\ref{eq:nc}) originates from the expression of the refractive index of a collisionless, unmagnetized plasma ${\sf n}(\omega)=\left(1-\omega_p^2/\omega^2\right)^{1/2}=\left(1-n_e/n_c\right)^{1/2}$ where $\omega_p=(4\pi e^2n_e/m_e)^{1/2}$ is the plasma frequency. Plasmas with density $n_e>n_c$ ($n_e<n_c$) are called overdense (underdense) and are opaque (transparent) to the laser light. For practical reasons it is useful to write $n_c$ as a function of the laser wavelength $\lambda_L=2\pi c/\omega_L$,  
\begin{equation}
n_c=\frac{\pi m_ec^2}{e^2\lambda_L^2}=\frac{1.1 \times 10^{21}~\mbox{cm}^{-3}}{(\lambda_L/1~\mu\mbox{m})^2} \; .
\label{eq:ncprat}
\end{equation}
This expression makes clear that for optical or near-infrared lasers with $\lambda_L \simeq 1~\mu\mbox{m}$ the cut-off density is about one hundredth (or less) of the electron density of solid targets. 

The second parameter is the dimensionless amplitude of the laser $a_0$, which corresponds to the oscillation momentum in the electric field of the laser in units of $m_ec$, i.e.
\begin{equation}
a_0=\frac{eE_L}{m_ec\omega_L}
=\left(\frac{e^2I_L\lambda_L^2}{\pi m_e^2c^5}\right)^{1/2}
=\left(\frac{I_L}{m_ec^3n_c}\right)^{1/2} \; ,
\label{eq:a0}
\end{equation}
where $\omega_L=2\pi c/\lambda_L$, $E_L$ and $I_L=cE_L^2/4\pi$ are the laser frequency, electric field amplitude, and intensity, respectively. A practical formula for $a_0$ as a function of $I_L$ and $\lambda_L$ is given by
\begin{equation}
a_0=0.85 \left(\frac{I_L\lambda^2_L}{10^{18}~\mbox{W cm}^{-2}\mu\mbox{m}^2}\right)^{1/2} \; .
\label{eq:a0-2}
\end{equation}
When $a_0>1$, the electron dynamics in the laser field is relativistic. Most of the experiments on ion acceleration have been performed with optical or near-infrared lasers and in the intensity range $I_L=10^{18}-10^{21}~\mbox{W cm}^{-2}$, which corresponds to $a_0 \simeq 0.85-28$ for $\lambda_L=1~\mu\mbox{m}$. CO$_2$ lasers with $\lambda_L \simeq 10~\mu\mbox{m}$  have been also used with typical intensities of  $I_L=10^{16}~\mbox{W cm}^{-2}$, yielding a mildly relativistic interaction regime.

The transmission of a laser pulse through a plasma is modified by relativistic effects on electron motion, which favor pulse penetration at densities higher than $n_c$, a phenomenon known as ``{relativistic transparency}''. Details depend on the laser and target parameters. A simple, although far from rigorous criterion applicable to targets much thicker than the laser wavelength $\lambda_L$ consists in assuming an increase of the cut-off density from $n_c$  to $n_c\gamma$ with $\gamma=(1+a_0^2/2)^{1/2}$, which is equivalent to assuming an effective electron mass equal to $m_e\gamma$ due the oscillation energy of electrons in the laser field. Notice that in general the effective mass will be a function of time and position since it will depend on the local amplitude of the electromagnetic field. This dependence produces a class of nonlinear optical effects, such as self-focusing and channeling of the laser pulse (see e.g. \citet{macchi-book}, chap.3). In addition, boundary effects are crucial for relativistic transparency: at the laser-plasma interface, the radiation pressure of the laser pulse pushes the plasma electrons and produces a local increase of the electron density, which counteracts the relativistic effect. As a result the transition to transparency has a higher threshold than would be obtained by simply assuming $n_e=\gamma n_c$ (see e.g. \citet{cattaniPRE00} and \citet{macchi-book}, chap.3.4.1).

For a target thinner than $\lambda_L$, the transparency threshold also depends on the target thickness $\ell$. As a case of particular importance for the following, the nonlinear reflectivity of a thin foil taking relativistic effects into account can be calculated in the limit of a Dirac-delta density profile (\citet{vshivkovPP98}; see also \citet{macchi-book}, sec. 3.4.2). The onset of  {relativistic transparency} is found to occur for
\begin{equation}
a_0 > \zeta \equiv \pi \frac{n_e\ell}{n_c\lambda_L} \; .
\label{eq:a0eqZeta}
\end{equation}

The generation of populations of ``suprathermal'' or ``fast'' electrons, i.e. populations of electrons having energy higher than the average energy of the bulk electrons, is of key importance. A theoretical analysis of fast electron generation mechanisms is outside the scope of the present paper (the reader may consult, e.g. \citet{gibbon-book,macchi-book} for introductory tutorials), thus we only resume the main features. Typically, fast electrons have broad energy spectra reasonably described with an exponential function $f({\cal E}_f)\propto \exp(-{\cal E}_f/T_f)$, extended up to a cut-off of a few times the ``temperature'' $T_f$ (energy units are used). The latter is often assumed to be of the order of the so-called ``ponderomotive energy'' ${\cal E}_p$, i.e.  the oscillation energy in the electric field of the laser:
\begin{equation}
{\cal E}_p = m_ec^2 \left(\left(1+a_0^2/2\right)^{1/2}-1\right) \; .
\label{eq:Th} 
\end{equation}
In current experiments on ion acceleration the range of laser intensity roughly corresponds to $a_0=1-30$, so that $T_f \simeq 0.1-10$~MeV may be expected (note, however, that measurements of $T_f$ are difficult, so that there is not a strong experimental background for the $T_f={\cal E}_p$ assumption).

For short laser pulses and sharp boundary targets (a situation typical of the interaction of femtosecond, high contrast pulses with solid targets) a simple picture of fast electron generation which supports Eq.(\ref{eq:Th}) is based on the so-called ``vacuum heating'' (VH) mechanism: the component of the Lorentz force perpendicular to the target surface periodically pulls electrons from the plasma into the vacuum side and then pushes them back into the plasma with an energy of the order of ${\cal E}_p$. In the non-relativistic regime ($a_0<1$) the VH mechanism is efficient at oblique laser incidence and for $P$-polarization: a simple model of the VH absorption coefficient ${\cal A}_f$ (i.e. the fraction of laser energy which is converted into fast electron energy via VH)) in the ${\cal A}_f\ll 1$ limits yields ${\cal A}_f\simeq a_0\sin^3\theta/\cos\theta$, where $\theta$ is the angle of incidence. In the $a_0>1$ regime, the expression for ${\cal A}_f$ is modified accounting for relativistic electron energies and laser pulse depletion (\citet{gibbon-book}, sec.5.5.2), showing that the dependence of ${\cal A}_f$ on $\theta$ and $a_0$ becomes weaker than for $a_0<1$. In addition, in the relativistic regime the magnetic part ($-e{\bf v}\times{\bf B}$) of the Lorentz force plays an important role and leads to efficient absorption and fast electron generation also for normal incidence. However, fast electron generation can be strongly quenched for normal incidence and circular polarization \citep{macchiPRL05} since in such conditions there is no component of the Lorentz force perpendicular to the target surface, suppressing the VH effect.

Assuming that the absorption is mostly due to fast electrons, a balance condition between the absorbed laser intensity and the flux of fast electron energy through the target may be written in order to estimate the density of fast electrons ($n_f$):
\begin{equation}
{\cal A}_f I_L \simeq n_f\upsilon_fT_f \; ,
\label{eq:balance}
\end{equation}
where $\upsilon_f$ is the fast electron velocity. In the non-relativistic regime, $\upsilon_f \simeq (2T_f/m_e)^{1/2}$ while $\upsilon_f \simeq c$ in the strongly relativistic case. Commonly, empirical values of ${\cal A}_f$ (typically of the order of $10\%$) and the ``ponderomotive'' estimate $T_f \simeq {\cal E}_p$ are used in Eq.(\ref{eq:balance}) to evaluate $n_f$. The values obtained are typically equal to a fraction of the cut-off density $n_c$, e.g. $n_f \sim 10^{20}~\mbox{cm}^{-3}$ for $\lambda_L\simeq 1~\mu\mbox{m}$. We remark again that these numbers should be considered only as gross estimates.

Note that both Eq.(\ref{eq:Th}) and the VH absorption coefficient ${\cal A}_f$ only depend on the laser intensity and not on target parameters. Actually, absorption and fast electron generation can be enhanced in targets which are weakly overdense, i.e. the electron density $n_e$ is at most a few times $n_c$. A rough explanation is that for such densities the laser frequency $\omega_L$ gets closer to the plasma frequency $\omega_p$ and thus the plasma response becomes more ``resonant'' (see \citet{mulserPRL08} and \citet{mulser-book}, sec.8.3.3 for a theoretical picture of absorption based on nonlinear, anharmonic plasma resonance). Moreover, the laser pulse may penetrate deeply into the plasma, leading to a stronger coupling. Eventually the low density favors the onset of relativistic transparency which is usually correlated with strong, turbulent heating of electrons; attempts to exploit this regime for ion acceleration are described in section~\ref{sec:SIT}.

Absorption and fast electron generation in solid targets also turn out to be sensitive to sub-wavelength density gradients and structuring of the interaction surface. Therefore, the pulse contrast (see sec.\ref{sec:shortpulse}) plays an important part. In particular, high contrast systems allow the use of suitable micro- and nano-structured targets to enhance and optimize ion acceleration, as will be discussed in sec.\ref{sec:structured}.

\subsection{Target Normal Sheath Acceleration}
\label{sec:TNSA}
The interpretation of the acceleration mechanism underlying the early observation of protons from solid targets \citep{clarkPRL00,maksimchukPRL00,snavelyPRL00} was the subject of some debate. Ultimately, the most successful description was that proposed by the LLNL group \citep{snavelyPRL00,wilksPoP01}, which was the basis of the so-called TNSA model. It is widely recognized that TNSA was the dominant mechanism for proton acceleration in most experiments reported so far (see \citet{macchiRMP13}, sec.III, for a detailed overview and list of references), including both the ``long'' and ``short'' pulse experiments included in Fig.\ref{fig:psdata} and Fig.\ref{fig:fsdata}.

\begin{figure}[t!]
\centering
\includegraphics[width=0.45\textwidth]{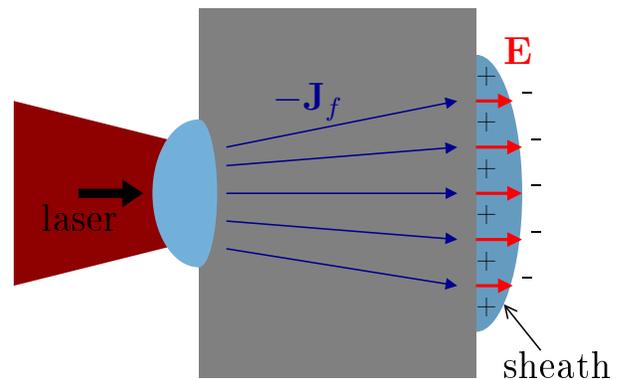}
\caption{The basic scheme of TNSA. The laser pulse incident on the front side of the target generates an intense current ${\bf J}_f$ or ``fast'' electrons which propagate through the target and produce a sheath at the rear side. The induced space-charge electric field ${\bf E}$ accelerates ions. Protons, which may be present either as a component of the target material or in hydrogen impurities present in a thin surface layer, are favored by the high charge-to-mass ratio and are thus preferentially accelerated.}
\label{fig:TNSAscheme}  
\end{figure}

TNSA is based on the efficient generation of fast electrons (Section~\ref{sec:scenario}) in the relativistic regime ($a_0>1$ or $I_L\lambda^2_L>10^{18}~\mbox{W}~\mu\mbox{m}^2\mbox{cm}^{-2}$). Fast electron generation produces very intense electrical currents into the target: the current density may be of the order of $J_f \simeq en_fc \sim 10^{11}~\mbox{A cm}^{-2}$ which may correspond to more than $10^5$~A through the focal area. If the target is relatively thin (from a few tens of microns down to sub-micrometric values) the fast electrons reach the rear side of the target (opposite to the laser-plasma interaction side, see Fig.\ref{fig:TNSAscheme}) producing a sheath region. In the sheath, a space-charge electric field is generated with a back-holding effect for the fast electrons which implies that the electric potential drop through the sheath is $\Delta\Phi \simeq T_f/e$. The field accelerates ions in the direction normal to the target surface. In metallic targets, protons are ordinarily present as surface impurities and are thus located near the peak of the sheath field; such localization, combined with the high charge-to-mass ratio, favors their acceleration with respect to heavier ions, unless the hydrogen containing layers are carefully removed \citep{hegelichPRL02,hegelichN06}. A test proton crossing the sheath region will acquire an energy ${\cal E}=e\Delta\Phi \simeq T_f$ which provides a first rough estimate of the energy gain and, assuming $T_f \simeq {\cal E}_p$ (Eq.\ref{eq:Th}), of the scaling of the proton energy with the pulse intensity.

Larger energy values may be achieved in the course of the expansion of the sheath plasma, where ultimately electrons and ions will reach the same drift velocity. In order to describe the acceleration mechanism several theory papers have revisited the classic problem of plasma expansion into vacuum in order to provide estimates for the proton energy as a function of the laser and target parameters (see e.g. \citep{kieferPRE13,moraPRL03,bettiPPCF05,moraPRE05,huangPoP13}, and references therein). Experimentally, the expanding sheath has been visualized by using the proton imaging technique, i.e. using a probe proton beam (also generated via TNSA) directed transversely to the plasma expansion \citep{romagnaniPRL05}. 

Establishing the scaling of the energy per nucleon ${\cal E}_n$ with laser and target parameters, and in particularly with the laser pulse energy ($U_L$) and intensity ($I_L$) is of fundamental importance to evaluate the potential of TNSA-based schemes for applications and give directions for further developments. 
Scaling laws have been inferred both by reviewing data from different laboratories (see e.g. \citet{borghesiFST06,borghesiPPCF08,daidoRPP12,kieferPRE13}) and by performing parametric studies on a single laser system (see e.g. \citet{fuchsNP06,robsonNP07,zeilNJP10}. In both cases, theoretical and semi-empirical models have been compared to the data. Inferred power-law scalings (${\cal E}_n \propto U_L^{\alpha}$ or ${\cal E}_n \propto I_L^{\alpha}$), with $\alpha$ ranging from 1/3 to 1 and possibly depending on the pulse duration, have been proposed. As we already discussed in sec.\ref{sec:state}, the uncertainty in establishing scaling laws might be ascribed both to the difficulties in the control and characterization of experimental conditions and to the several unknown quantities in models (such quantities are often used as parameters for fitting of experimental data). 

\subsection{Radiation Pressure Acceleration}
\label{subsec:RPA}\label{sec:RPA}
The incidence of an EM wave of intensity $I_L$ on a plane target leads to absorption of EM momentum, producing a pressure (for normal incidence)
\begin{equation}
P_{\mbox{\tiny rad}}=(1+{\cal R}-{\cal T})\frac{I_L}{c}=(2{\cal R}+{\cal A})\frac{I_L}{c} \; ,
\end{equation}
where ${\cal R}$, ${\cal T}$ and ${\cal A}$ are the reflection, transmission and absorption coefficients, respectively (energy conservation imposes the constraint ${\cal R}+{\cal T}=1-{\cal A}$). A maximum pressure $P_{\mbox{\tiny rad}}=2I_L/c$ is obtained in the case of an ideal ``perfect'' mirror with ${\cal R}=1$ and ${\cal T}={\cal A}=0$.
  
\begin{figure}[t!]
\begin{center}
\includegraphics[width=0.45\textwidth]{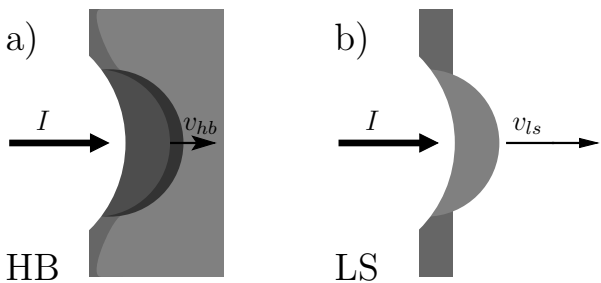}
\end{center}
\caption{Radiation pressure acceleration (RPA): simple illustration of the hole boring (HB) regime for thick targets (frame a) and of the light sail (LS) regime for thin targets (frame b).}
\label{fig:RPAscheme}
\end{figure}

In ultraintense laser interactions with overdense plasmas, the radiation pressure may overcome the thermal pressure and push as a piston the plasma steepening of the density profile and driving the recession on the interaction surface (Fig.\ref{fig:RPAscheme}~a). In a multi-dimensional geometry, such radiation pressure action bores a hole in the plasma, so that the velocity of the surface is commonly known as the ``{hole boring}'' (HB) velocity $u_{\mbox{\tiny HB}}$. Assuming steady conditions, the balance between the flows of electromagnetic and kinetic momentum at the surface yields $u_{\mbox{\tiny HB}}=(I_L/\rho c)^{1/2}$ (valid for $u_{\mbox{\tiny HB}}\ll c$; see \citet{robinsonPPCF09a} for a relativistic expression), where $\rho$ is the mass density. Moreover, the balance of mass and momentum flows at the moving piston surface implies that there must exist a flow of ions ``reflected'' from the recession front at twice $u_{\mbox{\tiny HB}}$, resulting in a ion population with energy per nucleon
\begin{equation}
{\cal E}_{\mbox{\tiny HB}}=\frac{m_p}{2}(2u_{\mbox{\tiny HB}})^2=\frac{2m_pI_L}{\rho c}=2m_ec^2\frac{Zn_c}{An_e}a_0^2 \; ,
\label{eq:Ehb}
\end{equation}
where we used $\rho \simeq Am_pn_i=(A/Z)m_pn_e$, being $Z$ and $A$ the ion charge and mass numbers, respectively. 
Eq.(\ref{eq:Ehb}) holds for a totally reflecting, cold plasma. Fast electron generation will reduce the HB efficiency by both decreasing ${\cal R}$ and producing a strong kinetic pressure which counteracts the radiation pressure. The quenching of fast electron generation by using circularly polarized pulses at normal incidence (section \ref{sec:scenario}) favors the radiation pressure action. The suppression of fast electrons also leads to reducing the intensity and energy of hard X-ray emission from the target, as experimentally observed \citep{aurandAPB15}, so that a ``cleaner'' source of energetic ions without secondary emissions may be obtained.

The scaling of ${\cal E}_{\mbox{\tiny HB}}$ with the density implies that for values typical of solid targets ($n_e>100n_c$) modest energies may be obtained.
Higher energies may be obtained via HB acceleration if the target density is reduced down to values slightly exceeding the cut-off density $n_c$ (lower density values are not suitable since the laser pulse would be transmitted through the target without any ``piston'' action). Combining such low-density targets with laser pulses at foreseeable intensities $I_L>10^{22}~\mbox{W cm}^{-2}$ may allow reaching $>100$~MeV energies as investigated theoretically \citep{macchiNIMA10,robinsonPoP11,robinsonPPCF12}.

Experimental evidence of HB acceleration has been provided in an experiment at the Accelerator Test Facility of Brookhaven National Laboratory (BNL) using a CO$_2$ infrared laser ($\lambda \simeq 10~\mu\mbox{m}$), for which $n_c \simeq 10^{19}~\mbox{cm}^{-3}$, and a hydrogen jet target having $n_e \gtrsim n_c$. Using pulses with circular polarization and  $I_L \simeq 10^{16}~\mbox{W cm}^{-2}$  proton spectral peaks at energy ${\cal E} \simeq 1~\mbox{MeV}$ have been observed \citep{palmerPRL11}. Recently, in a similar experiment the HB velocity has been measured using interferometric techniques to map in time the plasma profile \citep{gongPRE16}; it is found that the HB dynamics is affected by the peculiar temporal structure of the CO$_2$ pulse, which is actually a train of 3~ps ``micropulses'' separated by some tens of ps.

Using the long wavelength CO$_2$ laser as a driver enables the use of a flowing gas target, which may simplify high repetition rate operation compared to a solid target since the latter needs to be mechanically replaced or displaced in a very short time. However, in order to reach the ion energy range suitable for medical applications, the CO$_2$ laser intensity should be increased by two orders of magnitude; development projects and required technical advances are discussed by \citet{haberbergerOE10,bravyOE12,pogorelskyNIMA16}. 

Optical and near-IR lasers ($\lambda \simeq 0.8 - 1~\mu\mbox{m}$) presently produce much higher values of $a_0$ than CO$_2$ lasers. In order to maximize the energy obtainable via HB acceleration, a crucial step is the development of special targets with density values slightly above $n_c \simeq 10^{21}~\mbox{cm}^{-3}$, i.e. intermediate between those typical of gas jets and solid targets.  Possible approaches include very high density gas jets \citep{syllaRSI12} and foam materials with low average density \citep{zaniC13}. The production of a low-density ``preplasma'' by a laser prepulse impinging on solid targets before the main short pulse might also be a suitable strategy. Signatures of HB acceleration in solid targets have been associated with the interferometric observation of collimated plasma jets in petawatt interactions with foil targets of few micron thickness \citep{karPRL08b,karPPCF13}. 

If the target thickness $\ell\ll u_{\mbox{\tiny HB}}\tau_p$, with $\tau_p$ the laser pulse duration, the HB front reaches the rear side of the target before the end of the pulse duration; the whole mass of the target is then accelerated (strictly speaking the central region, as in Fig.\ref{fig:RPAscheme}~b), and the acceleration process can be iterated obtaining high velocities. This regime can be modeled, in its simplest form, as a thin mirror boosted by radiation pressure, i.e. a ``{light sail}'' (LS). Laser-driven sails were first proposed as a mechanism for interstellar travel propulsion \citep{marxN66,forwardJS84}, a concept recently brought back in press headlines because of the ``breakthrough starshot'' project \citep{meraliS16}.

The 1D model of a mirror composed by a single species (with charge and mass numbers $Z$ and $A$, respectively) having areal mass density $\rho\ell$ and boosted by a plane wave pulse of intensity $I_L$ and duration $\tau_p$ gives for the energy per nucleon (see e.g. \citet{macchi-book}, sec. 5.7.3)
\begin{eqnarray}
{\cal E}_{\mbox{\tiny LS}} = m_pc^2 \frac{{\cal F}^2}{2(1+{\cal F})} \; , 
\nonumber \\ 
{\cal F}=\frac{2I_L\tau_p}{\rho\ell c^2}=2\frac{Zn_c}{An_e}\frac{m_e}{m_p}\frac{\tau_pc}{\ell}a_0^2 \; .
\label{eq:LS}
\end{eqnarray}

Since ${\cal F}$ is inversely proportional to the areal density $n_e\ell$, using the thinnest targets as possible is advantageous in order to increase the particle energy. 
However, this approach is limited by the onset of pulse transmission through the target, which reduces the radiation pressure boost. Assuming the foil to be much thinner than the laser wavelength, the threshold for pulse transmission due to relativistic transparency effects is given by Eq.(\ref{eq:a0eqZeta}).
Interestingly, a relation equivalent to Eq.(\ref{eq:a0eqZeta}) is also obtained when the total radiation pressure is equal to the maximum electrostatic tension that can back-hold electrons in the foil; for higher intensities, the electrons are pushed away from the foil, and the ions undergo a Coulomb explosion, i.e. they are accelerated from their own space charge field which is unscreened by electrons. 
From both points of view, the condition (\ref{eq:a0eqZeta}) represents a compromise between maximizing the boosting radiation pressure on the foil and minimizing the foil mass, so it may be considered as an optimal working point for LS acceleration. It should be kept in mind, however, that  Eq.(\ref{eq:a0eqZeta}) is based on a simple model and plane geometry; in a realistic multi-dimensional geometry the expansion of the target in the transverse direction may lead to an earlier transition to transparency, limiting the energy gain. However, for high velocities of the target the reflectivity increases dynamically due to the increase of the laser wavelength in the frame co-moving with the foil, and (at least for relativistic ion velocities) the transverse expansion can reduce the areal mass on axis allowing to increase the maximum energy \citep{bulanovPRL10,sgattoniAPL14}. After these remarks, we still use Eq.(\ref{eq:a0eqZeta}) for a simple estimate of the energy gain in LS acceleration.
 
Inserting the optimal condition $a_0=\zeta$ from Eq.(\ref{eq:a0eqZeta}) in Eq.(\ref{eq:LS}) leads to an effective scaling (for non-relativistic ions i.e. ${\cal E}_{\mbox{\tiny LS}}=m_pV^2/2=m_pc^2{\cal F}^2/2$, ${\cal F}\ll 1$)
\begin{eqnarray}
{\cal E}^{(\mbox{\tiny opt})}_{\mbox{\tiny LS}} 
&=& 2\pi^2 m_pc^2\left(\frac{Z}{A}\frac{m_e}{m_p}\frac{c\tau_p}{\lambda}a_0\right)^2 \nonumber \\
&=& 2\pi^2 m_ec^2\left(\frac{m_e}{m_p}\right)\left(\frac{Z}{A}\frac{c\tau_p}{\lambda}a_0\right)^2 
\; .
\label{eq:LSopt}
\end{eqnarray}
For currently reachable laser intensities, the optimal thickness condition $a_0=\zeta$ requires $\ell \sim 10^{-2}\lambda$ or smaller for solid densities, that corresponds to nm-thick targets. Such ultrathin foils can be nowadays produced using, e.g. diamond-like carbon foil technology \citep{maNIMA11}. At the same time systems producing laser pulses with ultrahigh pulse contrast, such that no significant target ionization and damage is produced before the short pulse interaction, have been developed. Such combination of target and laser technology enables the experimental investigation of RPA-LS, which is appealing because of the favorable scaling with laser parameters (especially for sub-relativistic ion energies), the expectation of monoenergetic spectra (as all ions in the sail should move coherently with the same velocity), and the remarkable efficiency which comes with high sail velocity: in fact the mirror model predicts a degree of conversion of laser energy into pulse energy equal to $\eta=2\beta/(1+\beta)$ where $\beta$ is the sail velocity normalized to $c$, so that $\eta \simeq 40\%$ ($\beta \simeq 0.2$) for 100~MeV/nucleon ions. Such energies appear to be within reach with current laser and target technology: using a Ti:Sa laser ($\lambda=0.8~\mu\mbox{m}$) delivering 40~fs pulses ($c\tau_p/\lambda=15$) at an intensity of $10^{21}~\mbox{W cm}^{-2}$ ($a_0=22$), Eq.(\ref{eq:LSopt}) gives ${\cal E}^{(\mbox{\tiny opt})}_{\mbox{\tiny LS}} \simeq 150~\mbox{MeV}$.

At this point it is worth recalling that RPA-LS acceleration with ultraintense lasers was first proposed in 2004 on the basis of 3D simulations \citep{esirkepovPRL04} showing that the acceleration of thin targets at intensities exceeding $10^{23}~\mbox{W cm}^{-2}$ was well described by simple LS formulas. The extremely high intensity, still beyond current experimental capabilities, was considered to be necessary in order to enforce the dominance of RPA over other mechanisms (basically, the ions need to become relativistic within a laser cycle). However, later work showed that RPA is dominant at ``any'' intensity when circularly polarized pulses at normal incidence are used \citep{macchiPRL05}, which allows us to investigate LS acceleration using available laser systems as was proposed in theory papers \citep{zhangPP07,klimoPRSTAB08,robinsonNJP08}. Later work also suggested a dominance of RPA for linearly polarized pulses at intensities of $\sim 10^{21}~\mbox{W cm}^{-2}$ \citep{qiaoPRL12,macchiHPL14}. Despite these findings, it is still often quoted that RPA requires extreme intensities.

At first, LS seems not particularly suitable for proton acceleration: while a thin foil of solid hydrogen appears hardly feasible, in a multispecies target one expects that all components should ultimately move with the same velocity and thus the same energy per nucleon. This is based on the following argument: if light ions overcame heavier ones, the trailing edge of the sail would screen the laser pulse at their location, stopping the acceleration by radiation pressure. While LS would remain in any case a preferred option for the acceleration of multiply charged ions, an analysis of the LS dynamics beyond the rigid mirror model shows that, for pulses of finite duration, only part of the target ions are accelerated as a monoenergetic bunch, and in proper conditions this part may contain only the target protons \citep{macchiPRL09}; in addition, the formation of a region of accelerating field ahead of the target related to ``leaking'' transmission of the laser pulse may also accelerate protons \citep{qiaoPRL10}.

First experimental investigations of the RPA-LS regime \citep{karPRL12,henigPRL09a,palmerPRL12,dollarPRL12,aurandNJP13,steinkePRSTAB13} showed some promising results but also several issues, such as non-monoenergetic spectra, weak dependence on polarization, and non-uniformity of the accelerated beam. 
Indications of the transition to the LS regime, but also of non-optimized conditions, have been found in using the VULCAN {petawatt laser} ($I_L=0.5-3 \times 10^{20}~\mbox{W cm}^{-2}$, $\tau_p=0.7-0.9~\mbox{ps}$ and thin metallic targets ($\ell=0.1-0.8~\mu\mbox{m}$) containing carbon and hydrogen impurities \citep{karPRL12}. Narrow-band spectra (with energy spread $\sim 20\%$) were observed  centered at energies per nucleon approaching 10~MeV, displaying a scaling with $I_L\tau_p/\rho$ broadly consistent with the LS model prediction (\ref{eq:LS}).
The weak dependence on polarization is probably due to the relatively long laser pulse which lead to strong deformation of the target so that the incidence is not strictly normal anymore. 

More recently, an experiment performed on the GEMINI laser with a shorter pulse  ( $\tau_p=45~\mbox{fs}$, $I_L=6 \times 10^{20}~\mbox{W cm}^{-2}$) and ultrathin ($\ell=0.01-0.1~\mu\mbox{m}$) amorphous Carbon targets has given evidence of much higher Carbon ion energies for circular polarization (a 25~MeV cut-off energy to be compared with 10~MeV for linear polarization) \citep{scullionPRL17}. The analysis of the experiment, supported by 3D simulations, suggested that the energy gain was mostly limited by the onset of target transparency.

\subsection{Collisionless Shock Acceleration}
\label{subsec:shock}\label{sec:shock}\label{sec:CSA}
Under certain conditions, high intensity laser-plasma interactions lead to the generation of {collisionless shock waves}, i.e. sharp fronts of density and electric field which propagate in the plasma with supersonic velocity $V_s=Mc_s$, where the Mach number $M>1$ and $c_s=(ZT_e/Am_p)^{1/2}$ is the ``speed of sound'' (velocity of ion-acoustic waves) in a plasma. The term ``collisionless'' originates from the fact that, contrary to standard hydrodynamics, collisional and viscosity effects are not needed for the formation of the shock front, which is sustained by charge separation effects. 

\begin{figure}[t!]
\begin{center}
\includegraphics[width=0.45\textwidth]{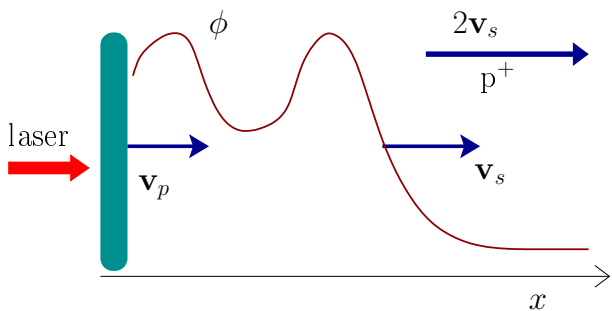}
\end{center}
\caption{Collisionless shock acceleration (CSA): simple illustration of the driving of a shock wave by the piston action of the radiation pressure of the laser, and of the reflection of protons from the shock front at twice the shock velocity $v_s$. The $\phi$ curve is a sketch of the electric potential associated to the electrostatic, collisionless shock \citep{tidman-krall}.}
\label{fig:CSA}
\end{figure}

The shock waves may be generated due to the piston action of the laser, i.e. the HB process (sec.\ref{subsec:RPA}) in a hot plasma, so that the density perturbation produced by radiation pressure detaches from the interaction surface and propagates in the plasma (Fig.\ref{fig:CSA}). This requires the shock velocity to be higher than the HB velocity $u_{\mbox{\tiny hb}}$, which may occur if the shock wave is sustained by the fast electron population. In such case we may replace the bulk electron temperature $T_e$ with the fast electron temperature $T_f$ in the expression for $V_s$. As an alternative mechanism, the fast electron stream might be subject to instabilities which ultimately develop shock waves. 

The electric field at the shock front may act as a potential barrier for ions in the plasma, accelerating some ions ``by reflection'' up to velocities $2V_s$: this is the basis for collisionless shock acceleration (CSA). The resulting energy per nucleon ${\cal E}_{\mbox{\tiny SA}}$ is (assuming a non-relativistic shock velocity)
\begin{equation}
{\cal E}_{\mbox{\tiny SA}}=\frac{m_p}{2}(2V_s)^2=2\frac{Z}{A}M^2T_f \; .
\end{equation}
As far as the shock front propagates at constant velocity, the reflected ions are monoenergetic.

CSA has been invoked as the mechanism leading to the generation of highly monoenergetic proton spectra (with up to $\sim 20~\mbox{MeV}$ peak energy) in an experiment performed using the Neptune CO$_2$ laser and gaseous hydrogen jet targets at the University of California at Los Angeles (UCLA) \citep{haberbergerNP12}. The energy spread of less than $1\%$ is the narrowest one observed in laser-plasma acceleration experiments. In this experiment, the pulse train structure (a sequence of 3~ps micropulses with peak intensity $I \simeq 6 \times 10^{16}~\mbox{W cm}^{−2}$) was found to be crucial to obtain the observed spectra: the density profile modification and plasma heating produced by the first micropulses in the train creates conditions favorable for the shock to be driven by the following micropulses. This suggests that a controlled, reproducible pulse sequence may be used to optimize CSA. Simulations of this scheme at higher intensities \citep{haberbergerNP12} suggested that CSA driven by CO$_2$ pulses in hydrogen gas jets could produce $>200~\mbox{MeV}$ protons at intensities of the order of $10^{18}~\mbox{W cm}^{-2}$, which would require a 20 times higher intensity compared to the currently most advanced CO$_2$ laser system.

So far, tailoring of the density profile for CSA optimization has been performed using a low intensity prepulse in experiments at BNL \citep{chenSPIE15,trescaPRL15}. With this approach, quasi-monoenergetic proton spectra with peak energy $\sim 1.2$~MeV have been obtained with CO$_2$ pulses at an intensity of $2.5 \times 10^{16}~\mbox{W cm}^{−2}$.  

The observed number of accelerated protons in the UCLA experiment \citep{haberbergerNP12} is very low, about three orders of magnitude lower than produced via HB acceleration at BNL (see sec.\ref{subsec:RPA}) in similar laser and target conditions \citep{palmerPRL11}. The comparison between the two experiments suggests that, for the same laser and plasma parameters, HB leads to lower energies than CSA but also to higher numbers of accelerated protons, which can be advantageous for some applications such as isochoric heating and creation of warm dense matter. In CSA, the number of accelerated protons must be low in order to prevent excessive ``loading'' of the shock wave: if too many protons are reflected from the shock, the latter loses energy and progressively reduces its velocity which in turn causes the reflected proton energy to shift down to lower values, broadening the proton spectrum \citep{macchiPRE12}. 

The control of the density profile is also crucial for the demonstration of CSA with optical lasers, which allow much higher intensities but also requires higher densities. A recent parametric study based on 3D simulations predicts that a proper combination of two laser pulses may allow CSA to produce $>100$~MeV protons using petawatt power systems \citep{stockemSR16}.

\subsection{Relativistic Transparency and Other Mechanisms}
\label{sec:underdense}\label{sec:SIT}
Not all the observations of ion acceleration may be fully explained in terms of TNSA, RPA or CSA. Depending on the laser and target parameters, the mechanism may be of hybrid nature combining aspects of all the three ``basic'' acceleration concepts. In addition, in the literature there are many proposals of particular schemes, employing e.g. complex target configurations, and which are typically supported by numerical simulations. Only a minority of such proposals have been investigated in the laboratory so far. For the sake of brevity in this section we restrict us to experimental observations and to the main trends in hybrid or alternative schemes. 

As discussed in sec.\ref{sec:RPA}, the onset of relativistic transparency in thin targets is apparently the main factor limiting the RPA-LS scheme. However, a different approach to ion acceleration actually exploits transparency to reach high energies. The basic idea is that the transition from opacity to transparency is accompanied by an efficient, although somewhat turbulent energy absorption by the plasma electrons (see sec.\ref{sec:scenario}), which may couple to ions also via collective plasma modes or instabilities. This regime has been investigated in particular at the TRIDENT laser facility at Los Alamos National Laboratory (LANL) \citep{henigPRL09a,hegelichNF11,jungNJP13,jungNJP13b,hegelichNJP13}. In latest reported experiments at LANL, C and Al ions were accelerated up to 18~MeV/nucleon with up to 5\% conversion efficiency and energy spread as narrow as 7\% for laser pulses of up to 80~J energy, 650~fs duration, and intensity up to $8\times 10^{20}~\mbox{W cm}^{-2}$ \citep{palaniyappanNC15}. 

Ion acceleration using foil targets undergoing transparency has been studied recently also using the PHELIX laser at GSI Darmstadt \citep{wagnerPoP15} and at VULCAN \citep{powellNJP15}, both laser systems having pulse duration of several hundreds of fs like TRIDENT. These experiments suggest a complex acceleration scenario, where different mechanisms (either TNSA-like or RPA-like) contribute at different stages and produce typical signatures in the energy spectrum and angular distribution of the ions. In some conditions the cut-off energies achieved in this regime may be higher than with respect to TNSA from thicker targets at similar laser parameters, but typically such energies are observed at some angle with respect to the laser axis. 

The transparency regime is also accessible with ultrashort (tens of fs) pulses if ultrathin (tens of nm) targets are used. Experiments in this regime performed with the PULSER I laser at the Advanced Photonics Research Institute (APRI) of GIST, Korea \citep{kimPRL13} and at MBI Berlin \citep{braenzelPRL15} have identified signatures of Coulomb explosion, which occurs over the transparency threshold (see sec.\ref{sec:RPA}). In particular, for the MBI experiments this regime seems favorable for efficient acceleration of heavy (Gold) ions \citep{braenzelPRL15}.
Ultrathin targets also enabled a different exploitation of relativistic transparency in controlling proton acceleration which has been recently explored in an experiment on the GEMINI laser \citep{gonzalez-izquierdoNC16}: in the transparent regime the proton spatial distribution is sensitive to the laser pulse polarization, which may be then used to control the proton beam.

An alternative approach to generate a near-transparent, or near-critical ($n_e \simeq n_c$) plasma for high absorption is to use special target material such as a foam. An experiment on VULCAN investigated relativistic transparency and related proton acceleration using foam targets \citep{willingalePRL09}. Experiments on foam-covered foils for enhanced in TNSA have been performed on PULSER \citep{prencipePPCF16,passoniPRAB16} and will be further discussed in sec.\ref{sec:structured}.

A few experiments employed underdense ($n_e<n_c$) plasmas produced from gas jets, which would be advantageous for high repetition rate operations (such as gas jets used in combination with CO$_2$ lasers for CSA, see sec.\ref{sec:CSA}). In this regime, an intense laser pulse may generate a charge-displacement channel with an electrostatic field accelerating ions mostly in the direction radial with respect to the propagation axis. However, He ions collimated along the propagation direction were also observed using VULCAN \citep{willingalePRL06} with a 40~MeV cut-off at $6\times 10^{20}~\mbox{W cm}^{-2}$ intensity and 1~ps duration, and also using the JLITE-X laser at JAEA-KPSI at much lower intensity ($7\times 10^{17}~\mbox{W cm}^{-2}$ intensity, with 40~fs duration) with a surprising 20~MeV cut-off \citep{fukudaPRL09}. The interpretation of these experiments has stimulated the proposal of a particular mechanism, named Magnetic Vortex Acceleration (MVA) where the accelerating electric field is generated by magnetic induction via the formation of electron vortices at the plasma-vacuum interface \citep{bulanovPRL07}.

\section{Advanced Optimization Strategies}
\subsection{Target Engineering}
\label{sec:structured}
Since its formulation the TNSA model has provided a framework for optimization of ion acceleration by manipulating the target properties. 
Manipulation of the chemical composition on the rear surface has been used for spectral manipulation. The effect is related to the expansion of a multi-species plasmas, where depending on the relative concentration between heavy and light ions (typically protons) peaks are formed in the spectrum of the light species (see e.g. \citet{tikhonchukPPCF05} and sec.III.C.r of \citet{macchiRMP13}). In addition, concentration of hydrogen-containing molecules in thin dots coated on the rear side was exploited in order to reduce spectral broadening due to the transverse inhomogeneity of the sheath field, obtaining spectral peaks with $\sim 10\%$ spread \citep{schwoererN06,pfotenhauerNJP08}. As a specular approach, removal of hydrogen impurities was used for TNSA of heavier species, e.g. Carbon \citep{hegelichPRL02}, also obtaining spectral peaks in particular conditions \citep{hegelichN06}. So far, however, these approaches have been limited to modest ion energies (a few MeV per nucleon) and progress appears to have been slow.

\begin{figure*}[th!]
\begin{center}
\includegraphics[width=0.8\textwidth]{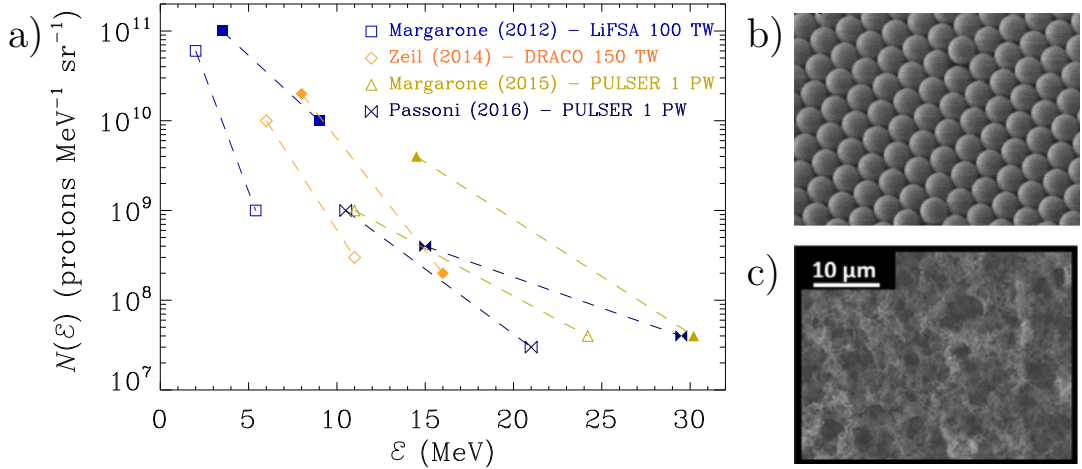}
\end{center}
\caption{a): proton energy spectra from experiments comparing flat and structured targets. The flat target data are already included in Fig.\ref{fig:fsdata} and shown here with empty symbols: filled symbols corresponds to the same experimental parameters, but using structured targets. These latter included thin foils covered by sub-micron size spheres \citep{margaronePRL12}, mass-limited targets \citep{zeilPPCF14}, and foam-covered foils \citep{passoniPRAB16}. b): scanning electron microscope (SEM) image of a layer of $0.94~\mu\mbox{m}$; adapted with permission from \citet{floquetJAP13}. c): SEM image of  a Carbon foam; adapted with permission from  \citet{passoniPPCF14}. }\label{fig:fsdataST}
\end{figure*}

Engineering of the front side of the target has been oriented to increase the energy and conversion efficiency of fast electrons, in order to obtain enhanced TNSA. As already mentioned in sec.\ref{sec:state}, focusing the laser pulse into gold microcones placed on the front surface of thin foils yielded a $\sim 30\%$ increase in the cut-off energy and about one order of magnitude increase in proton number with respect to the flat case in an experiment performed at LANL (Fig.\ref{fig:psdata}).

It has also been observed that a non-planar surface, with structures having a size of the order of the laser wavelength, may allow more efficient laser absorption and, in turn, higher proton cut-off energies \citep{margaronePRL12,floquetJAP13,margaronePRSTAB15,prencipePPCF16,passoniPRAB16}. Clearly, the exploitation of micro- or nano-structures requires the use of very high contrast, femtosecond pulses, otherwise the structuring would be washed out by the prepulse or, in any case, well before the peak pulse intensity.

Another strategy is to use ``mass-limited'' targets, i.e. using targets with a reduced volume to confine the absorbed energy and obtaining higher temperatures: for example, enhancement of conversion efficiency and proton cut-off energy has been observed in foils with limited (tens of microns) transverse extension \citep{buffechouxPRL10,zeilPPCF14}.  In some cases, isolated targets with no mechanical support such as droplets have been used \citep{ter-avetisyanPRL06,sokollikPRL09,ter-avetisyanPoP12,ostermayrPRE16}.

\begin{figure*}[t!]
\includegraphics[width=0.8\textwidth]{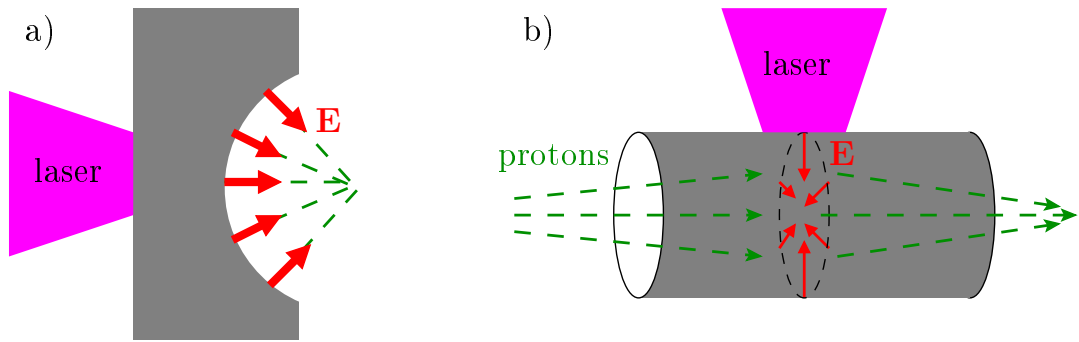}
\caption{a): ``static'' focusing of TNSA-produced protons by a spherical target. b): ``dynamic'' focusing of protons by the electric field generated inside a laser-irradiated cylinder \citep{toncianS06}.}
\label{fig:shaping}
\end{figure*}

For some of the experiments mentioned in sec.\ref{sec:shortpulse} with data shown in Fig.\ref{fig:fsdata}~a), the flat target results were compared with those obtained with engineered targets. These latter include foil targets with the irradiated surface either covered by micro- or nano-sphere layers (Fig.\ref{fig:fsdata}~b) \citep{margaronePRL12,margaronePRSTAB15} or by foam layers (Fig.\ref{fig:fsdata}~c) \citep{passoniPRAB16}, and transversely limited foils \citep{zeilPPCF14}. In all these cases, the comparison with the flat target was made at the same value of the foil thickness. Fig.\ref{fig:fsdataST} shows that structured targets may yield up to a $\sim 50\%$ enhancement of the energy cut-off and to more than one order of magnitude increase in the spectral density of protons at a given energy. The enhancement effect produced by the microsphere layer is further discussed by \citet{floquetJAP13}. Additional observations in the foam-covered target experiments \citep{passoniPRAB16}, such as independence on pulse polarization, suggest that the enhancement is due to a geometrical effect related to the microscopic structure of the foam which allows efficient volumetric heating. 

Special targets may be also designed in order to allow high absorption via the excitation of resonances in the laser-produced plasma (sec.\ref{sec:scenario}). 
For example, foam targets can be produced with average density $n_e \simeq n_c$ so that the laser field would couple with bulk plasma oscillations or plasmons. However, this interaction scenario probably requires some pre-heating of the foam target in order to smooth its inhomogeneous structure, which is preserved in high contrast interactions (see the discussion in the preceding paragraph). In ``grating'' targets with a periodic modulation at the laser-irradiated surface, the laser pulse can couple instead to surface plasmons at angles of incidence $\theta_i$ such that $\sin\theta_i \simeq 1-j\lambda/d$, where $d$ is the spatial period and $j$ is an integer number. In an experiment with the UHI laser (25~fs, 80TW) at the SLIC facility of CEA Saclay, high contrast pulse interaction with grating targets at angles around the $j=1$ resonance has shown to produce a factor of $~\sim 2.5$ increase in the proton cut-off energy with respect to flat targets at the same angle of incidence \citep{ceccottiPRL13}. 

Nanostructured targets may also be used with acceleration mechanisms different from TNSA. Recently, in order to investigate RPA-LS with extremely intense and sharp rising fs pulses, thin foil targets covered by a few-micron Carbon nanotube foam (CNF) on the interaction side have been used in order to generate self-focusing and self-steepening of the laser pulse in a plasma of density close to $n_c$. Using such technique on the GEMINI laser delivering 50~fs pulses at $I=2 \times 10^{20}~\mbox{W cm}^{-2}$, enhanced acceleration of carbon ions (up to $\sim$20~MeV energy per nucleon) with RPA features has been observed \citep{binPRL15}.

Very recently, first experiments have been performed employing cryogenic targets. Advantages of such technology include the low electron density (down to a few tens of $n_c$) and the suitability for high repetition rate since the target is flowing. In addition, use of cryogenic hydrogen affords obtaining pure proton spectra. Preliminary results have been obtained with the TITAN laser at LLNL \citep{gauthierRSI16} and with the PALS laser (600~ps duration) in Prague \citep{margaronePRX16}.

\subsection{Optical Control and Post-Acceleration}
\label{sec:postacceleration}

In principle, laser accelerated ions can be handled using conventional accelerator techniques for energy selection, transport, and focusing. However, it is of interest to investigate handling techniques which can be integrated in the laser and target configuration, i.e. on a very short and compact scale similar to that over which the acceleration occurs. 
Most of these techniques has been developed in the TNSA framework although they might be adapted to other laser acceleration schemes.

Already in the first experiments on proton acceleration it was observed that the direction of protons could be simply controlled by shaping and orientation of the rear surface of the target,  since protons are emitted in the direction normal to such surface. It was thus apparent that a focusing of the proton bunch could be obtained by using a target with spherically shaped rear surface (see Fig.\ref{fig:shaping}~a). This static approach has been used for the first time by \citet{patelPRL03} where proton focusing was exploited to isochorically heat matter.

Further developments of the target shaping concept have been investigated by \citet{karPRL08a,burzaNJP11,karPRL11,bartalNP12,chenPRL12}. Some of these target configurations were designed in order to exploit the transient nature of the TNSA field for dynamic focusing: since the field travels along the target as a unipolar surface wave with picosecond duration \citep{quinnPRL09,tokitaSR15}, the target can be designed in a way that the fields interacts at its peak with protons of a given energy, yielding energy selection capability. Alternatively, a double pulse scheme may be used as in the experiment of \citet{toncianS06} where a TNSA bunch was focused by the field  generated inside a hollow cylinder by a secondary pulse (Fig.\ref{fig:shaping}~b): by changing the synchronization between the two pulses, a particular energy slice can be focused and selected.

\begin{figure*}[t!]
\begin{center}
\includegraphics[width=0.8\textwidth]{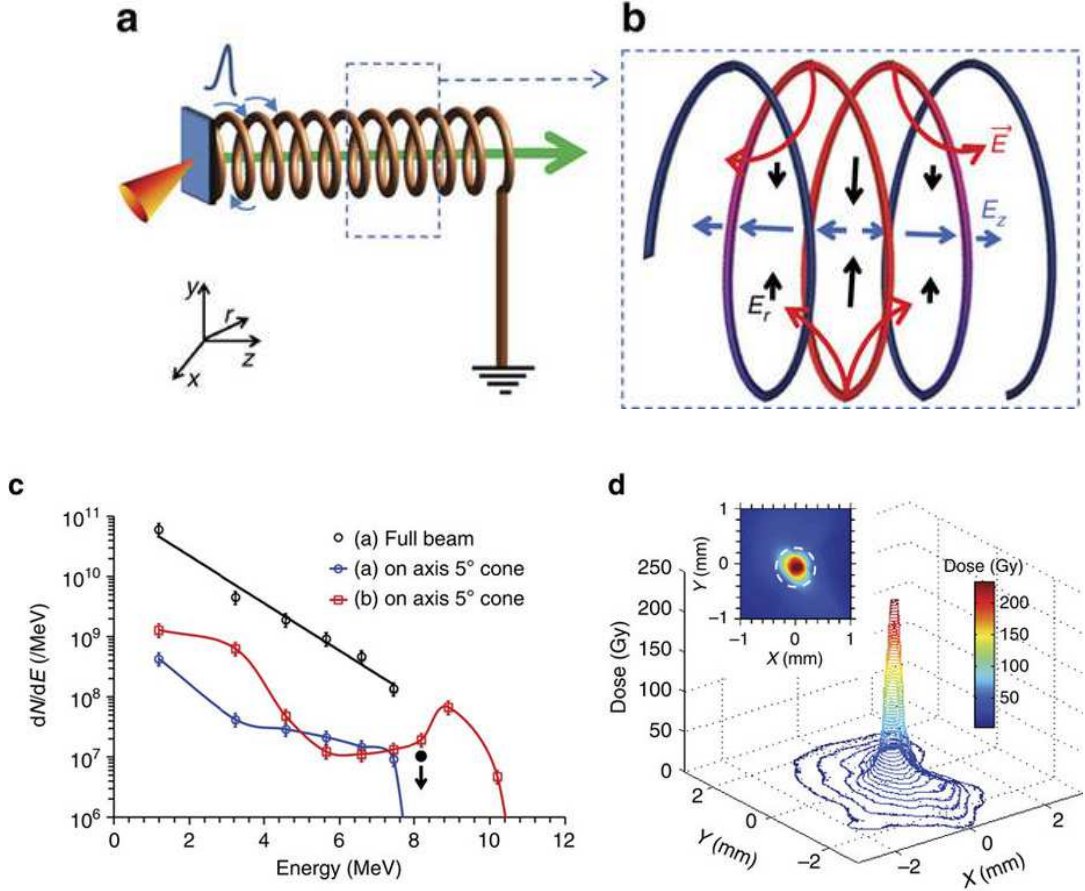}
\caption{Proton beam focusing and post-acceleration using an helical coil attached to the rear side of the target \citep{karNC16}. a): the scheme of the device. b): sketch of the electric field distribution inside the coil, in correspondance of the travelling signal. c): comparison of spectra with and without the coil, showing the energy increase and spectral bunching near cut-off for protons travelling near the axis. d): the narrow spatial distribution of 8.9~MeV protons after focusing by the coil. Adapted with permission from \citet{karNC16}.}
\label{fig:coilexp-2}
\end{center}
\end{figure*}

Recently, the above ideas have been further developed in a novel device for post-acceleration with chromatic focusing and energy enhancement capability \citep{karNC16,kar-patent}. The design of the device is shown in Fig.\ref{fig:coilexp-2}~a). A metallic wire attached to the rear side of the target is bent in order to form an helical coil coaxial with the proton beam. After the interaction, while the protons are accelerated in the sheath the unipolar pulse propagates along the coil, its velocity along the longitudinal direction being determined by the coil radius and pitch. This allows the unipolar pulse to be synchronized with protons of a given velocity. The electric field generated near the axis of the coil has both radial components directed towards the axis, and longitudinal components which are both parallel or anti-parallel to the axis as shown in \ref{fig:coilexp-2}~b). Thus, a fraction of the protons may be simultaneously focused and further accelerated by the electric field of the traveling SW, as it has been demonstrated in proof-of-principle experiments \citep{karNC16} using the ARCTURUS laser system at Heinrich Heine University in D\"usseldorf. The main results are summarized in Fig.\ref{fig:coilexp-2}~c)-d): an increase of the cut-off energy by $\sim 2$~MeV is obtained  in combination with a tight focusing of the protons along the axis of the system.

\section{Discussion and Outlook}
\label{sec:outlook}\label{sec:conclusions}
As already mentioned in this chapter, evaluating the progress made in laser-driven ion acceleration requires a critical insight. For instance let us focus for the moment on the enhancement of the cut-off energy. While the latter has been considered the most relevant parameter to assess the progress in the field, its experimental determination may be ambiguous particularly for broad exponential-like spectra, as it depends on the detection method and its sensitivity and on the practices for signal discrimination from background noise. While establishing reliable and agreed experimental procedures remains a priority for the community, our analysis in sec.\ref{sec:state} has made an attempt to perform a meaningful comparison by taking full, calibrated spectra into account. 

The most relevant result emerging from such analysis is, in our opinion, the enhancement of the cut-off energy achieved with ``small'', short-pulse laser systems with a scaling law which appears more promising than inferred in previous work, as shown in Fig.\ref{fig:fsdata}; roughly, the scaling is linear in the pulse energy \emph{in the focal spot} with a $\sim$9~MeV/J slope. Establishing whether this scaling will be maintained for larger pulse energies, delivering such energies in the focal spot, and providing a theoretical support for this scaling are possible near-term research goals.

As already mentioned, two experiments in our selection \citep{kimPRL13,oguraOL12} have made claims of cut-off energies significantly higher than the values predicted from the scaling law at the same laser pulse energy. Possible explanations for these ``anomalous'' data might be searched in differences in the experimental set-up. 
For the experiment of \citet{oguraOL12}, where $0.8~\mu\mbox{m}$ thick targets were used, the only apparent difference with respect to other references is a lower pulse contrast, since plasma mirrors were not used; in principle this might favor the formation of a small preplasma  which may already affect the interaction. 

The experiment of \citet{kimPRL13} was performed on the PULSER laser using ultrathin targets (in the 10-100~nm range) and intensities up to $3.3 \times 10^{20}~\mbox{W cm}^{-2}$ which might lead to an acceleration regime different from TNSA, with strong effects of both RPA (sec.\ref{sec:RPA}) and transparency (sec.\ref{sec:underdense}). A hybrid acceleration mechanism with overlapping contributions by both TNSA and RPA was also proposed on the basis of a more recent experiment with the ATLAS laser system at MPI Garching ($I_L=8 \times 10^{19}~\mbox{W cm}^{-2}$, $\tau_L=30~\mbox{fs}$), also using ultrathin targets ($5-20~\mbox{nm}$).
measured the proton cut-off energy ${\cal E}_{\mbox{\tiny co}}$ as a function of the absorption coefficent ${\cal A}$, obtaining a scaling ${\cal E}_{\mbox{\tiny co}}\propto{\cal A}$ in agreement with models predicting a linear scaling of ${\cal E}_{\mbox{\tiny co}}$ with the absorbed laser energy \cite{schreiberPRL06,zeilNJP10}. The extremely thin targets used ($5-20~\mbox{nm}$) favored a contribution of RPA (section \ref{sec:RPA}) to the proton energy, so that the acceleration mechanism may be considered of hybrid nature.

Hybrid regimes of accelerations (sec.\ref{sec:underdense}) are possibly promising for proton acceleration but still require thorough investigations before more difficult applications can be tackled on their basis.  
Similarly, mechanisms such as CSA (sec.\ref{sec:CSA}) or magnetic vortex acceleration (sec.\ref{sec:underdense}) are still in a very early stage of investigation and more experiments are needed both to confirm preliminary findings and to confirm theoretical predictions.

Recently, a proton energy cut-off of 93~MeV has been reported from another campaign on PULSER \citep{kimPoP16} using the thinnest targets (10~nm) employed by \citet{kimPRL13}. At the highest intensity ($7 \times 10^{20}~\mbox{W cm}^{-2}$) only, the cut-off energy was larger for circular polarization than for linear polarization, so that efficient RPA-LS acceleration (sec.\ref{sec:RPA}) was claimed although the proton spectrum was broad (with no clear spectral peak) and anomalously modulated. The accuracy of determining the maximum energy (strongly at variance with experiments performed for similar parameters in different laboratories) was at the limit of the detector (Thomson parabola) range with possible issues of low resolution, noise floor and trace discrimination. For these reasons further support by additional experiments will be required. An expansion of the data on RPA-LS is expected with growing number of PetaWatt-laser systems starting their operation, since the RPA-LS mechanism remains very attractive because of its scaling properties; the control and understanding of the onset of transparency and the target stability are expected challenges on the route to RPA optimization.

Enhancement of the energy per nucleon is by no means the only required development to make laser-driven ion acceleration suitable for applications. For instance, while at least for ``large'' laser drivers (Fig.\ref{fig:psdata}) the cut-off energy now falls in the therapeutic window for ion beam therapy in oncology\footnote{Ocular tumors may be treated with $\sim$60~MeV protons, however most hadrontherapy treatments require 200-225~MeV protons at the surface of the patient}, laser-driven acceleration remains quite far from the other stringent requirements \citep{linzPRSTAB07,linzPRAB16}. At present it appears not possible to predict either when such requirements will be reached or if such developments in laser-driven schemes will be faster than those in approaches based on conventional accelerators, which are becoming more compact and cheaper. Probably, the key issue will be whether the unique properties of laser-accelerated ions may be advantageously exploited for therapeutic benefits. For example, the avalaibility of extremely high dose rate could enable the first investigation of ``collective'' regimes in the biological response to irradiation \citep{fourkalPMB11}, and fast optical control of the short-duration ion bunches might be useful for the irradiation of moving organs \citep{hofmannJB12}.

While medical applications remain a long-term challenge, laser-accelerated protons have already had a major impact as a diagnostic of laser-plasma interactions, providing data which are also of broad interest for nonlinear science and astrophysically relevant phenomena. An increasing use of laser-accelerated protons and ions in production of warm dense matter and time-resolved studies of material damage is also likely (see chapter 9 by Dromey and chapter 16 by Nordlund). These applications exploit the short duration of the ion bunch and its natural synchronization with laser pulses, which open the way to pump-probe experiments.

We expect that such perspectives will stimulate further developments in laser-driven ion acceleration. 
Such developments may be supported both by the expected increase in available laser intensities and by progress in target engineering, including micro- and nano-structuring and the use of special materials. With the support of theory and massively parallel numerical simulation, hopefully such advances will allow us to reach significant milestones in the next decade.

\hyphenation{Post-Script Sprin-ger}

\end{document}